\documentclass[runningheads]{llncs}

\usepackage{eccv}

\usepackage{eccvabbrv}

\usepackage{graphicx}
\usepackage{booktabs}

\usepackage[accsupp]{axessibility}  % Improves PDF readability for those with disabilities.

\usepackage{hyperref}

\usepackage{algorithm}
\usepackage{algorithmic}
\usepackage{tabularx}
\usepackage{enumitem}
\usepackage{multirow}

\usepackage{orcidlink}

\begin{document}

\title{Spatially-Variant Degradation Model for Dataset-free Super-resolution} 

\titlerunning{Spatially-Variant Degradation Model for Dataset-free SR}

\author{Shaojie Guo\inst{1}\orcidlink{0009-0008-3955-7240} \and
Haofei Song\inst{1}\orcidlink{0009-0006-4926-7525} \and
Qingli Li\inst{1}\orcidlink{0000-0001-5063-8801} \and
Yan Wang\thanks{Corresponding author} \inst{1} \orcidlink{0000-0002-1592-9627}}

\authorrunning{Guo et al.}

\institute{East China Normal University, Shanghai 200241, China \\
\email{\{52275904013,hfsong\}@stu.ecnu.edu.cn},\email{ qlli@cs.ecnu.edu.cn},\email{ ywang@cee.ecnu.edu.cn}}

\maketitle

\begin{abstract}
This paper focuses on the dataset-free Blind Image Super-Resolution (BISR). 
Unlike existing dataset-free BISR methods that focus on obtaining a degradation kernel for the entire image, 
we are the first to explicitly design a spatially-variant degradation model for each pixel. Our method also benefits from having a significantly smaller number of learnable parameters compared to data-driven spatially-variant BISR methods. Concretely, each pixel's degradation kernel is expressed as a linear combination of a \textbf{learnable} dictionary composed of a small number of spatially-variant atom kernels. The coefficient matrices of the atom degradation kernels are derived using membership functions of fuzzy set theory. We construct a novel Probabilistic BISR model with tailored likelihood function and prior terms. Subsequently, we employ the Monte Carlo EM algorithm to infer the degradation kernels for each pixel. Our method achieves a significant improvement over other state-of-the-art BISR methods, with an average improvement of 1 dB ($2\times$).Code will be released at \url{https://github.com/shaojieguoECNU/SVDSR} 
\end{abstract}

\section{Introduction}
\label{sec:intro}

Super-Resolution (SR) aims to reconstruct high-resolution (HR) images from their degradation red, noisy low-resolution (LR) counterparts. 
Due to the challenge of obtaining the degradation  kernel in real-world scenarios, Blind Image Super-Resolution (BISR) has recently garnered significant research interest.
Mathematically, SR degradation model can be formulated as:
\begin{equation}
\boldsymbol{y}=(\mathcal{D} \boldsymbol{x}) {\downarrow_s} +\boldsymbol{n}, 
\label{eq:srmodel}
\end{equation}
where $\boldsymbol{x}$ denotes the HR image, $\boldsymbol{y}$ denotes the LR image, $\mathcal{D}$ denotes degradation operator,  $\downarrow_s$ denotes subsampling operator with scale factor $s$, and $\boldsymbol{n}$ denotes the noise. 

Numerous BISR methods have been proposed in recent years to reconstruct images afflicted by complex and unknown degradation operators.
These methods can be broadly classified into two categories. 
Methods\cite{kim2016accurate,lai2018fast,hui2019lightweight,lu2022transformer} belonging to the first category directly use complex deep neural networks (DNNs) to map the observed LR images to HR images, which do not effectively utilize the physical information (\emph{e.g.}, edge and texture) of image degradation in \cref{eq:srmodel} and may yield suboptimal results. 
Methods belonging to the second category first estimate the degradation operator $\mathcal{D}$ from the LR images and then transform the blind super-resolution problem into a non-blind SR problem. 
The challenge with this approach lies in how to accurately estimate the unknown degradation operator. 
Estimating the degradation operator directly from the LR image is a highly ill-posed problem.
Many priors for degradation operator have been proposed to better estimate the degradation kernel. \cite{yue2022blind,liang2021flow, ren2020neural, bell2019blind} designed explicit or implicit kernel priors for the entire image to better utilize the information within LR images, these priors are all spatially-invariant.
\cite{chen2023better,kim2021koalanet,liang2021mutual,zhou2023learning,li2020lapar} formulated a unique degradation kernel for each patch or even each pixel of the image. 
These spatially-variant adaptive degradation kernels can fully utilize the structural information differently for each pixel location, thereby obtaining richer visual details in the reconstructed image.
However, learning its own degenerate kernel for each pixel is computationally expensive.
\cite{zhou2023learning,li2020lapar} use a linear combination of 72 pre-defined atom kernels to represent the degenerate kernels of different pixels, which greatly reduces the computational burden.
They meticulously designed the networks to learn the huge coefficient matrix of the atom kernel for each pixel, rather than focusing on learning the shape of the entire degradation  kernel.
But these networks still need to be trained on large paired artificially generated dataset, which significantly reduces the practicality of the proposed methods.

This work introduces a novel spatially-variant degradation model for dataset-free BISR, which enjoys the benefit of (1) dealing with spatially-variant degradation model for dataset-free BISR and (2) much smaller numbers of learnable parameters than data-driven BISR.
Our approach enables the determination of individual degradation kernels for each pixel in the image without the requirement of a dataset. 
Each pixel's degradation kernel in our model is expressed as a linear combination of a dictionary of \textbf{learnable} atom kernels.
In accordance with the observation in \cite{li2020lapar}, the shape of the degradation kernel is highly correlated with the texture density of the area where its corresponding pixel is located.
Therefore, we innovatively use fuzzy set theory to derive the coefficient matrix of the atom kernels based on the texture information of the images. 
Different from the fixed atom kernel dictionary used in previous works, various degradation kernels are enabled to be more flexibly represented thanks to the learnability and adaptability of the proposed learnable dictionary.
Our approach eliminates the need to learn the shape of the entire atom kernels. Each atom kernel in the proposed dictionary is uniquely determined by only three learnable parameters, which are theoretically supported by high-dimensional Gaussian kernel decomposition technology \cite{keilmann2023improved}, significantly reducing the number of learning parameters.
Additionally, we meticulously design the likelihood function and priors under the Maximum A Posteriori (MAP) framework to further enhance the performance of the proposed model.

Our main contributions can be summarized as follows:
\begin{itemize}
\item The first dataset-free deep learning method for BISR with an explicit spatially-variant degradation model. 
To the best of our knowledge, all existing spatially-variant degradation models require extensive training on large datasets.
\item A novel Probabilistic BISR Model with elaborately designed likelihood function and prior terms.
An inference algorithm for the proposed model is also derived based on the Monte Carlo Expectation Maximization (MCEM) algorithm. 
\item Extensive experiments on synthetic datasets and real images verify the effectiveness of our method.
Compared to sota dataset-free BISR methods, there is an average improvement of 1 dB ($2\times$) achieved.
\end{itemize}

\section{Related Works}

\subsubsection{Non-learning BISR Methods:}
Most existing non-learning methods adopt an iterative approach to search the HR image.
These models are designed to incorporate various handcrafted image priors, such as sparsity priors \cite{rudin1992nonlinear,sun2008image,krishnan2009fast,kim2010single} and non-local similarity priors \cite{dong2012nonlocally,he2020non}, to address the ill-posed nature of the BISR problem.
However, these methods often yield poor performance due to the inherent structural complexity of natural images.

\subsubsection{Data-driven Deep Learning Methods:} 
Many deep learning methods \cite{fritsche2019frequency,he2019ode,ji2020real,kim2016accurate,liu2020residual,maeda2020unpaired,mei2020image,tai2017memnet,wang2018recovering,xie2019learning,zhang2018image,li2020lapar} have been proposed for BISR task. 
Here, we provide a brief introduction to spatially-variant degradation-based models.
Shocher \etal~ \cite{shocher2018zero} propose a Zero-Shot SR (ZSSR) model to leverage the recurring patterns of image patches across different scales to achieve SR.
Liang \etal~ \cite{liang2021mutual} partition images into patches and suggests a Mutual Affine Network (MANet) to emulate space-variant blur for each patch.
Kim \etal~ \cite{kim2021koalanet} design the Kernel-Oriented Adaptive Local Adjustment (KOALA) method for jointly learning spatially-variant degradation and restoration kernels.
Chen \etal~ \cite{chen2023better} design a Cross-MOdal fuSion network (CMOS) that estimate both blur and semantics simultaneously for images with out-of-focus blur. 
Zhou \etal~ \cite{zhou2023learning} improve SR results by adjusting degradation to known degradation using the proposed linearly-assembled pixel degradation-adaptive regression module (DARM).
\cite{zhou2023learning,li2020lapar} utilize a dictionary of multiple pre-defined filter bases to transform the complex degradation  kernel prediction task into a linear coefficient regression task.
\cite{wang2021real,zhang2021designing,mou2022metric} use the HR datasets and their degradation algorithm to generate the HR-LR dataset to train their model.
These methods all require training on large generated datasets and do not fully exploit physical priors for kernel inference.

\subsubsection{Dataset-free deep learning Methods:}  
Training a neural network for BISR without a dataset is challenging. 
Existing dataset-free methods all assume that the priors proposed are spatially-invariant.
Ulyanov \etal~ \cite{ulyanov2018deep} propose the Deep Image Prior (DIP) framework,
which leverages a generator network as an implicit prior for SR tasks without the need for a dataset.
Liang \etal~ \cite{liang2021flow} and Yue \etal~ \cite{yue2022blind} respectively introduce implicit flow-based kernel prior (FKP) and explicit kernel prior (EKP) into DIP to enhance the performance of BISR.
Our method also falls into this category. 
However, in contrast to previous approaches which all assume that the degradation kernel is spatially-invariant, 
we are a pioneer in proposing the spatially-invariant degradation model using deep learning methods without relying on datasets.
The unique degradation kernel for each pixel is established by leveraging the texture information in its local neighborhood. This approach tailors the degradation process for each pixel based on its individual characteristics.
Besides, the degeneration kernel in Yue et al.~\cite{yue2022blind} is assumed to follow a simple Gaussian distribution, whereas our model extremity enhances its representation through the utilization of a Gaussian mixture model.

\section{The Proposed Model}
\begin{figure*}[t]
  \centering
\includegraphics[width=1\linewidth]{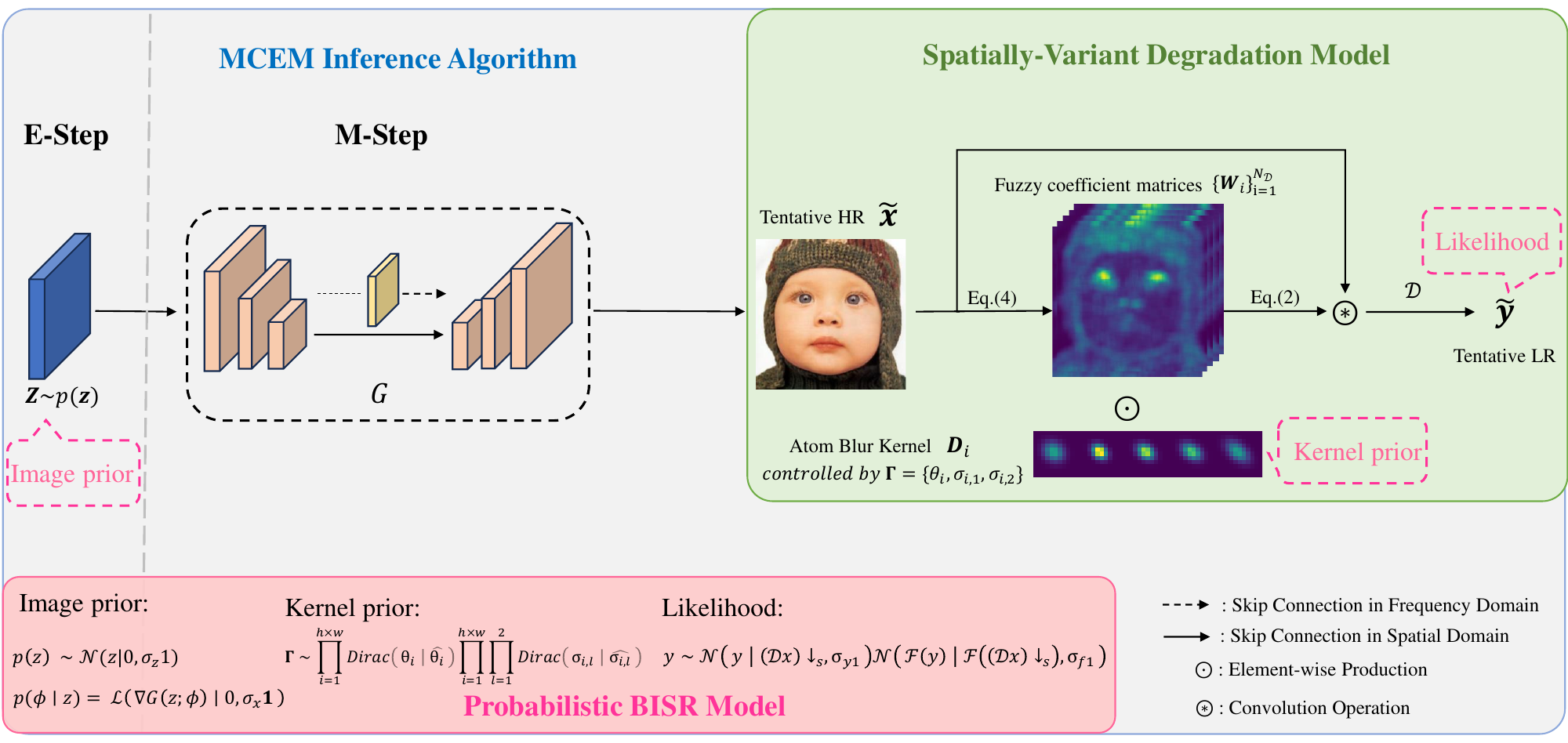}
\caption{The framework of the entire model and its solution (MCEM inference algorithm). The proposed Spatially-Variant Degradation Model consists of  of a dictionary of learnable atom operators $\{\boldsymbol{D}_i\}_{i=1}^{N_\mathcal{D}}$ 
and corresponding coefficient matrices $\{\boldsymbol{W}_i\}_{i=1}^{N_\mathcal{D}}$ .
Each atom degradation operator is determined by three learnable parameters $\theta_i$,  $\sigma_{i,1}$ and  $\sigma_{i,2}$. The coefficient matrices $\{\boldsymbol{W}_i\}_{i=1}^{N_\mathcal{D}}$ are obtained from the tentative HR image $\Tilde{\boldsymbol{x}}$. 
Image prior, kernel prior and likelihood are suggested to solve the proposed spatially-variant degradation model under the MAP framework. 
In the inference process, the latent variable $\boldsymbol{z}$ is sampled in the E-Step and parameters $\{\boldsymbol{\Gamma}\}_{i=1}^{N_{\mathcal{D}}}$ and weights $\phi$ of G are updated in M-step. }
\label{fig:framework}
\end{figure*}

In this section, we first suggest a spatially-variant degradation model with the HR image $\boldsymbol{x}$ and variables 
% $\boldsymbol{W}_{i}$ and $\mathcal{D}_{i}$ 
related to the degradation operator (see \cref{sec:SVDM}). Then, to model the variables in the degradation model, a probabilistic model is employed under the Maximum A Posteriori (MAP) framework, which incorporates uncertainty and probability distributions to represent and analyze the variables (see \cref{sec:PModel}). %and then we develop a probabilistic model to  based on the degradation model  (see \cref{sec:PModel}). 
In the next section, we will utilize the MCEM algorithm to solve the proposed probability model (see \cref{sec:mcem}). The entire model and its solution are shown in \cref{fig:framework}.

\subsection{Spatially-Variant Degradation Model}
\label{sec:SVDM}

Inferring the degradation kernel for each pixel is computationally expensive and challenging, especially in the absence of a dataset. Therefore, having reasonable prior assumptions is crucial for improving the computational efficiency and practicality of spatially-variant degradation models.
Considering that each pixel in the same photo undergoes a similar imaging process, a lower-dimensional manifold should exist in the degradation operator space of a whole photo.
We utilize the O'Leary \cite{nagy1998restoring} formula to decompose degradation operator as follows:
\begin{equation}
(\mathcal{D} \boldsymbol{x})[h, w]:=\sum_{r, c} \sum_{i=1}^{N_{\mathcal{D}}} \boldsymbol{W}_{i}[h, w]  \mathcal{D}_{i}  \boldsymbol{x}[h-r, w-c],
\label{eq:degradation operator}
\end{equation}
where $\{\mathcal{D}_{i}\}^{N_{\mathcal{D}}}_{i=1}$ is a learnable dictionary composed of a set of atom degradation  kernels,
$\{\boldsymbol{W}_{i}\}^{N_{\mathcal{D}}}_{i=1}$ is a set of coefficient matrices, $N_{\mathcal{D}}$ is the number of atom degradation  kernels,
$[h, w]$ denotes the location of each pixel, $r, c$ denotes the location of the surrounding pixels.
More than that, each element of the degradation  kernel should obey the physical constraints that it is non-negative and the sum of all elements is equal to one,
the constraints $\sum_{i=1}^{N_{\mathcal{D}}} \boldsymbol{W}_{i}[h, w] = 1$, $\boldsymbol{W}_{i} \geq 0$ are also imposed on \cref{eq:degradation operator}.
Note that in previous linearly-assembled degradation models\cite{li2020lapar}, the dictionary $\{\mathcal{D}_{i}\}^{N_{\mathcal{D}}}_{i=1}$ is pre-defined and the coefficient matrices $\{\boldsymbol{W}_{i}\}^{N_{\mathcal{D}}}_{i=1}$ need to be learned and its size is usually related to the size of the image.
We further model $\mathcal{D}_{i}$ and $\boldsymbol{W}_{i}$ separately.

\paragraph{Atom degradation  Operator $\mathcal{D}_{i}$:}
%\noindent {\bf Atom degradation  Operator $\mathcal{D}_{i}$:} 
Inspired by the the fact that any continuous distribution can be approximated with any specific non-zero amount of error by a Gaussian mixture model with sufficient components \cite{mclachlan2019finite}.
The atom degradation  operator $\mathcal{D}_{i}$ is expressed through convolution with an anisotropic Gaussian degradation  kernel $\boldsymbol{g}_{i}$ in \cref{eq:degradation operator}.
Additionally, the anisotropic Gaussian degradation  kernel can be decomposed \cite{keilmann2023improved} as follows:
\begin{equation}
\begin{aligned}
\mathcal{D}_{i} \triangleq \boldsymbol{g}_{i}\left(x_{1}, x_{2}\right)= 
 \prod_{l=1}^{2} \frac{1}{\sqrt{2 \pi} \sigma_{i,l}} \exp \left(- \frac{\left(\boldsymbol{x}^{\top} \boldsymbol{\nu}_{l}\right)^{2}}{2\sigma_{i,l}^{2}}\right),
\end{aligned}
\label{eq:gdecompose}
\end{equation}
 where $\boldsymbol{x}=\left({x}_{1},{x}_{2}\right)^{\top} \in \mathbb{R}^{2}$,
 $\boldsymbol{\nu}_{1}$$=(\cos (\theta_i), \sin (\theta_i))$ and  $\boldsymbol{\nu}_{2}$$=(-\sin (\theta_i), \cos (\theta_i))^{\top}$,
 $\sigma_{i,l}$ are the variances of the two isotropic Gaussian kernels obtained by decomposing $\boldsymbol{g}_{i}$, and $\theta_i \in  [0, \pi)$ is the rotation angle of the coordinate axis.
 Note that for the given $\boldsymbol{W}_{i}$, \cref{eq:degradation operator} is differentiable \emph{w.r.t.} the set of free parameters $\boldsymbol{\Gamma}=\{\theta_i,\sigma_{i,1},\sigma_{i,2}\}$,  indicating that $\{\mathcal{D}_{i}\}$ are learnable.
 The learnable atom degradation operator is in stark contrast to previous spatially-variant degradation models, where the atom degradation kernel is fixed.

\paragraph{Coefficient Matrices $\boldsymbol{W}_{i}$:}
The shape of the degradation kernel is significantly different between pixels in flat areas and high texture density areas\cite{li2020lapar}.
Inspired by the correlation between the shape of the degradation kernel and the texture richness of its region, we propose to estimate the set of coefficient matrices $\boldsymbol{W}_{i}$ for the atom degradation kernel through a sequence of membership functions of fuzzy set theory, utilizing the tentative HR image $\tilde{\boldsymbol{x}}$ as follows:

\begin{equation}
\boldsymbol{W}_{i} = \frac{\boldsymbol{\mu}_{i}(\tilde{\boldsymbol{x}})}{\sum_{i=1}^{N_{\mathcal{D}}} \boldsymbol{\mu}_{i}(\tilde{\boldsymbol{x}})}.
\label{eq:wi}
\end{equation}
 
\cref{eq:wi} is a sum normalization operation, aiming to ensure the physical prior that the sum of all elements is equal to one. 
As a reminder, when $N_{\mathcal{D}}=1$, the coefficient matrix corresponding to the only degradation kernel is an all-ones matrix, which makes the proposed degradation model naturally degenerates into a conventional spatially-invariant degradation model.
When $N_{\mathcal{D}} \ge 2$, membership function $\{\boldsymbol{\mu}_{i}\}^{N_{\mathcal{D}}}_{i=1}$ are defined as follows:
\begin{equation}
\boldsymbol{\mu}_{i}(\tilde{\boldsymbol{x}})= \text{exp} \left(- \frac{\left(N_{\mathcal{D}}-1\right)}{2\sigma_g^2}
\left(\boldsymbol{h}\left(\tilde{\boldsymbol{x}}\right)-\frac{i-1}{N_{\mathcal{D}}-1}\right)^2 \right),
\label{eq:mui}
\end{equation}
where $\sigma_g$ is the shape parameter. {$\boldsymbol{h}(\cdot)$} is the feature extraction function used to extract the texture features of $\tilde{\boldsymbol{x}}$, \emph{i.e.},
\begin{equation}
\boldsymbol{h}\left(\tilde{\boldsymbol{x}}\right) =\boldsymbol{H} * (\nabla \tilde{\boldsymbol{x}}),
\label{eq:ft}
\end{equation}
where $*$ denotes the convolution operator, $\boldsymbol{H}$ is a median filter with window size $P \times P$, $\nabla$ is a first-order derivative used to extract the texture features of $\tilde{\boldsymbol{x}}$. 

\noindent {\bf Remark 1.} 
The proposed degradation model offers a more compact yet expressive approach compared to previous spatially-invariant degradation models~\cite{yue2022blind,liang2021flow,liang2021mutual,kim2021koalanet,li2020lapar}, which fully utilizes the low dimensional manifold information in the imaging process and greatly reduces the number of learned parameters.
The learnable degradation kernel dictionary ensures the expressive ability of the proposed degradation model, and the coefficient matrix obtained through fuzzy set theory makes full use of the physical information in the image.
This allows the proposed spatially-variant degradation model to be effectively utilized on a single LR image without the need for extensive dataset training.
% For a given $512 \times 512$ image with a degradation  kernel size of $15 \times 15$, the ratio of the number of learned parameters of our model to MANet (patchsize: $9\times9$)\cite{liang2021mutual} is $3N_\mathcal{D}: 73,955$, {where $N_\mathcal{D}$ is usually a small number}.

\subsection{Probabilistic BISR Model}
\label{sec:PModel}
\noindent

We construct a probabilistic BISR model based on the proposed Spatially-Variant degradation model illustrated in Sec.~\ref{sec:SVDM}.
Under the MAP framework, \cref{eq:srmodel} can be modeled as a MAP estimation as follows:

\begin{equation}
\max _{\boldsymbol{D}, \boldsymbol{x}} \log p(\boldsymbol{y}\mid\boldsymbol{D}, \boldsymbol{x})+\log p(\boldsymbol{x})+\log p(\boldsymbol{D}),
\label{eq:map}
\end{equation}
where $p(\boldsymbol{y}\mid\boldsymbol{D},\boldsymbol{x})$ represents the likelihood, while $p(\boldsymbol{x})$ and $p(\boldsymbol{D})$ denote image prior and kernel prior. 
We elaborately design these three terms to achieve accurate HR reconstruction without relying on a dataset.

\paragraph{Likelihood:}
Unlike most existing methods that only consider the likelihood $p(\boldsymbol{y} \mid \boldsymbol{D}, \boldsymbol{x})$ in the spatial domain, in this work we define a likelihood in both the spatial domain and the frequency domain as:
\begin{equation}
 y \sim \mathcal{N} (\boldsymbol{y} \mid \left( \mathcal{D} \boldsymbol{x}\right)\boldsymbol{\downarrow}_s, \sigma_{y}\boldsymbol{1}) \mathcal{N} (\mathcal{F}\left(\boldsymbol{y}\right) \mid \mathcal{F}\left(\left( \mathcal{D} \boldsymbol{x}\right)\boldsymbol{\downarrow}_s\right) , \sigma_{f}\boldsymbol{1} ),
\label{eq:likelihood}
\end{equation}
where $\boldsymbol{1}$ represents the identity matrix, $\sigma_{y}$ and $\sigma_{f}$ are hyper-parameters which represent the variances of Gaussian white noise in the spatial and frequency domains, respectively.
The likelihood in the frequency domain enhances the reconstruction effectiveness of the proposed model.
An ablation study of the frequency domain likelihood can be found in \cref{sec:exp}.

\paragraph{Kernel Prior:} 
Based on the aforementioned implicit constraints priors and physical priors, a spatially-variant degradation model is plugged into \cref{eq:map} as the kernel prior. 
Next, each element of $\boldsymbol{\Gamma}$ can be naturally defined as Dirac distributions with the ground truth $\hat{\boldsymbol{\Gamma}}=\{\hat{\theta_i},\hat{\sigma}_{i,1},\hat{\sigma}_{i,2}\}$, \emph{i.e.},
\begin{equation}
\boldsymbol{\Gamma} \sim \prod_{i=1}^{h\times w}\operatorname{Dirac}(\theta_i \mid \hat{\theta_i}) \prod_{i=1}^{h\times w}\prod_{l=1}^{2} \operatorname{Dirac}(\sigma_{i,l} \mid \hat{\sigma}_{i,l}).
\label{eq:dirackernelprior}
\end{equation}

Since the ground truth ${\hat{\theta_i}}$ are unknowns in BISR, we adopt Gaussian distributions with zero mean and variance $\sigma_\gamma$ to represent \cref{eq:dirackernelprior} as:
\begin{equation}
\boldsymbol{\Gamma} \sim \prod_{i=1}^{h\times w}\mathcal{N}(\theta_i \mid 0,\sigma_\gamma) \prod_{i=1}^{h\times w}\prod_{l=1}^{2} \mathcal{N}(\sigma_{i,l} \mid 0,\sigma_\gamma).
\label{eq:gaussiankernelprior}
\end{equation}

\noindent {\bf Remark 2.} 
By employing \cref{eq:dirackernelprior} and \cref{eq:gaussiankernelprior}, our model is versatile, capable of accommodating both blind SR with Dirac distributions using the ground truth $\hat{\boldsymbol{\Gamma}}$ and non-blind SR with Gaussian distributions featuring a zero mean and variance $\sigma_\gamma$.

\paragraph{Image Prior:} 
We approximately characterize the complex structural nature of natural images by reparametrizing the latent space using a network $G$ with 3-layer encoder-decoder structure U-net, \emph{i.e.},
\begin{equation}
\boldsymbol{x} = G\left(\boldsymbol{z};\boldsymbol{\phi}\right),
\label{eq:netprior}
\end{equation}
where $\boldsymbol{z}$ denotes the latent variable, $\boldsymbol{\phi}$ denotes the deterministic weights of $G$.  
The joint distribution of $\boldsymbol{\phi}$ and $\boldsymbol{z}$ is defined as:
\begin{align}
(\boldsymbol{\phi},\boldsymbol{z}) \sim p(\boldsymbol{\phi},\boldsymbol{z}) = 
p(\boldsymbol{\phi} \mid \boldsymbol{z})p(\boldsymbol{z}).
\label{eq:jointprior}
\end{align}

Considering the powerful fitting capability of CNN, conventional Laplacian prior with scale $\sigma_x$ and Gaussian prior with variance $\sigma_z$ are to mitigate the overfitting phenomenon. These priors are defined as:
\begin{align}
p(\boldsymbol{\phi} \mid \boldsymbol{z}) & = 
\mathcal{L}(\nabla G\left(\boldsymbol{z};\boldsymbol{\phi}\right) \mid 0, \sigma_x \boldsymbol{1}), \label{eq:imageprior1} \\
p(\boldsymbol{z}) & \sim \mathcal{N}(\boldsymbol{z} \mid 0, \sigma_z \boldsymbol{1}). \label{eq:imageprior2}
\end{align}
where $\boldsymbol{z}$ denotes the first coefficient matrix used to emphasize texture information.
Additionally, we incorporate instance normalization layers on $G$ and skip connection structure in the frequency domain like \cite{mao2023intriguing} to alleviate overfitting in a dataset-free setting.

\section{MCEM Inference Algorithm}
\label{sec:mcem}

\begin{algorithm}[t]
	%\textsl{}\setstretch{1.8}
	\renewcommand{\algorithmicrequire}{\textbf{Input:}}
	\renewcommand{\algorithmicensure}{\textbf{Output:}}
	\caption{MCEM Inference Algorithm}
	\label{alg1}
	\begin{algorithmic}[1]
            \REQUIRE Observed LR image $\boldsymbol{y}$, hyper-paramter settings.
            \ENSURE  the estimated HR image $\tilde{\boldsymbol{x}}$.
		\WHILE  {not converged}
		\STATE $\textbf{E-Step:}$  Sample the latent variable $\boldsymbol{z}$ from \cref{eq:ldsample}
		\STATE $\textbf{M-Step}$ Update parameter $\boldsymbol{\Gamma}$ and $\boldsymbol{\phi}$ based on \cref{eq:minproblem}
		\ENDWHILE
		\STATE   $\tilde{\boldsymbol{x}}=\boldsymbol{G}(\boldsymbol{z};\phi)$
	\end{algorithmic}  
\end{algorithm}
In this section, we utilize the MCEM inference algorithm to solve the proposed probabilistic BISR model in Sec.~\ref{sec:PModel}.
First, we remodel BISR according to the preview equation. 
Then, in the E-step, we adopt Langevin dynamics (LD)
 as a Monte Carlo sampler to  search for the optimal distribution of $\boldsymbol{z}$ and maximize the Evidence Lower Bound (ELBO) with respect to the model parameters in the M-step.
According to \cref{eq:map,eq:likelihood,eq:dirackernelprior,eq:gaussiankernelprior,eq:netprior,eq:jointprior,eq:imageprior1,eq:imageprior2},
we re-model BISR with an additional inference of parameters:
\begin{equation}
\begin{aligned}
\underset{\boldsymbol{\Gamma}, \boldsymbol{\phi}}{\max }  \ & p(\boldsymbol{\Gamma},  \boldsymbol{\phi} \mid \boldsymbol{y}) =  \log \int  p(\boldsymbol{y} \mid \boldsymbol{\Gamma}, \boldsymbol{\phi}, \boldsymbol{z} ) p(\boldsymbol{\Gamma} \mid \boldsymbol{y}) p(\boldsymbol{\phi} \mid \boldsymbol{z})p(\boldsymbol{z}) \mathrm{d} \boldsymbol{z}+\text {const}.
\end{aligned}
\label{eq:logproblem}
\end{equation}

\paragraph{E-Step:}
In the E-Step, we fix the parameters in \cref{eq:logproblem} and search for the optimal distribution of $\boldsymbol{z}$. 
We denote it as $p(\boldsymbol{z} \mid \boldsymbol{y}, \boldsymbol{\Gamma}_\text{old}, \boldsymbol{\phi}_\text{old})$, where $\{\boldsymbol{\Gamma}_\text{old}, \boldsymbol{\phi}_\text{old}\}$ represents the current parameters.
$\boldsymbol{z}$ can be approximately sampled using Stochastic Gradient Langevin Dynamics (SGLD) for $\tau = 1, 2, \ldots, n_z$.

\begin{equation}
\boldsymbol{z}^{\tau+1}=\boldsymbol{z}^{\tau}+\alpha \frac{\partial}{\partial \boldsymbol{z}} \log p(\boldsymbol{z}^{\tau} \mid \boldsymbol{y}, 
\boldsymbol{\Gamma}_{old}, \boldsymbol{\phi}_{old})
+\sqrt{2\alpha}\boldsymbol{\zeta},
\label{eq:ldsample}
\end{equation}
where $\alpha$ denotes the step size satisfying Robbins-Monro condition, $\zeta\sim\mathcal{N}(0,\boldsymbol{1})$ represents Gaussian white noise used to prevent trapping into local modes.

The posterior of $\boldsymbol{z}$ can be calculated according to \cref{eq:likelihood,eq:imageprior1,eq:imageprior2} as:

\begin{equation}
\begin{aligned}
p(\boldsymbol{z} \mid \boldsymbol{y},   \boldsymbol{\Gamma}_\text{old}, \boldsymbol{\phi}_\text{old}) & = \frac{1}{2{\sigma}_y^2}\left\| \boldsymbol{y}-\mathcal{D}_\text{old} \boldsymbol{x}_\text{old}\right\|_{2}^{2} 
+\frac{1}{2\sigma_x} \left\| \nabla \boldsymbol{x}_\text{old}\right\|\\
& +\frac{1}{2{\sigma}_f^2}\left\| \mathcal{F}\left(\boldsymbol{y}\right)-
\mathcal{F}\left(\mathcal{D}_\text{old} \boldsymbol{x}_\text{old} \right)
\right\|_{2}^{2}
+\frac{1}{2\sigma_z^2}\left\|\boldsymbol{z}\right\|_{2}^{2},
\label{eq:gz}
\end{aligned}
\end{equation}
where $\mathcal{D}_\text{old}$ denotes the degradation  operator determined by $\boldsymbol{\Gamma}_\text{old}$, and $\boldsymbol{x}_\text{old}$ denotes the HR image generated by $G\left(\boldsymbol{z};\boldsymbol{\phi}_\text{old}\right)$.

\paragraph{M-Step:}
The M-step aims to maximize the Evidence Lower Bound (ELBO) \emph{w.r.t.} the model parameters.
Fixing the sampled latent variable $\boldsymbol{z}$ in the E-step,  the objective function for optimizing model parameters $\boldsymbol{\Gamma}$ and $\boldsymbol{\phi}$ can be formulated based on \cref{eq:likelihood,eq:gaussiankernelprior,eq:imageprior1} as:

 \begin{equation}
\begin{aligned}
 \max _{\boldsymbol{\Gamma}, \boldsymbol{\phi}} Q(\boldsymbol{\Gamma},  \boldsymbol{\phi}) 
 & \Rightarrow  \min _{\boldsymbol{\Gamma}, \boldsymbol{\phi}}  
 E(\boldsymbol{\Gamma}, \boldsymbol{\phi}) \ \frac{a\ln(1+\boldsymbol{W}^{down}_{N_D})+b}{2{\sigma_y}^2}\left\| \boldsymbol{y}-\mathcal{D} \boldsymbol{x} \right\|_{2}^{2}+\frac{1}{2{\sigma_\gamma}^2} \sum_{i=1}^{h\times w}
 \left\|\theta_i\right\|_{2}^{2} \\
 & +\frac{1}{2{\sigma_f}^2}\left\| \mathcal{F}\left(\boldsymbol{y}\right)-
\mathcal{F}\left(\mathcal{D}\boldsymbol{x} \right)
\right\|_{2}^{2}+\frac{1}{2\sigma_x} \left\| \nabla \boldsymbol{x}\right\|
 +\frac{1}{2{\sigma_\gamma}^2}\sum_{i=1}^{h\times w}\sum_{l=1}^{2}\left\|\sigma_{i,l}\right\|_{2}^{2}.
\end{aligned}
\label{eq:minproblem}
\end{equation}
where a and b are empirically set to 45000 and 8000.$\boldsymbol{W}^{down}_{N_D}$ denotes the last layer coefficient matrix after sampling. \cref{eq:minproblem} can be solved via ADAM optimizer. 
The whole MCEM Inference Algorithm is illustrated in \cref{alg1}.

\section{Experiments}
\noindent

In the rest of this paper, we refer to our Spatially-Variant Degradation Model for dataset-free Super-Resolution as SVDSR. 

\label{sec:exp}
\subsection{Experimental Settings}
\paragraph{Parameter Settings:}
Throughout the experiments, we set the number of atom degradation  kernels $N_\mathcal{D}$ as 5. 
For the membership functions, a median filter with a square window size of 15 is utilized to extract the texture information from the image, and the shape parameter of the membership function $\sigma_g$ is empirically set to 0.5.
For simplicity, the variance $\sigma_y$ and $\sigma_f$ of the spatial domain and frequency domain in likelihood are set to 1 and 2, respectively. 
A generation network $G$ with 3-layer encoder-decoder structure U-net is employed to predict the HR image,
and the Laplace distribution prior with a variance $\sigma_x$ of 2.5 and a Gaussian distribution prior with a variance $\sigma_z$ of 1 are used to alleviate the occurrence of overfitting.
Since the real degradation  kernels are unknown in the BISR task, we empirically use a Gaussian distribution with a variance $\sigma_\gamma$ of 1.5 to simulate their distribution. 
The hyperparameters $\alpha$ and $n_z$ for $\boldsymbol{z}$ sampling are set to 1.5 and 5.
For the sake of fairness in comparison, the learning rates for $\gamma$ and $\phi$ are set to $2 \times 10^{-3}$ and $5 \times 10^{-3}$, respectively, which are the same as BSRDM \cite{yue2022blind}. 
The maximum iteration number of the EM algorithm is capped at 5,000.

\paragraph{Comparison Methods:}
To evaluate our model, we compare it against five methods, namely RCAN~\cite{zhang2018image}, ZSSR~\cite{shocher2018zero}, DoubleDIP~\cite{li2023self}, DIPFKP~\cite{liang2021flow} and  BSRDM~\cite{yue2022blind}. 
RCAN  is a  residual channel attention networks with residual in residual (RIR) structure.
LARPAR\cite{li2020lapar} is not used for comparison due to its degradation kernel and downsampling methods are different compared to common dataset-free methods.
ZSSR is a zero-shot method that exploits the patch recurrence within the same image scale, as well as across different scales in a single image.
Since the degradation models of ZSSR and RCAN are both assumed to be a bicubic downsampler, it is unfair to compare with their methods on datasets generated using Gaussian degradation  kernels.
In addition to testing the above methods in the BISR, we also test the effect of ZSSR in the non-blind case. 
The ground truth of the degradation  kernel is provided by us.
By incorporating the non-blind method, different models can be better evaluated.
DoubleDIP, DIPFKP, and BSRDM are the same type of comparison methods as our method, which propose different degradation models to better reconstruct image details.
Among them, the degradation model of BSRDM is spatially-invariant and BSRDM is known as a state-of-the-art (SOTA) method in Dataset-free BISR.
All the code used for comparison comes from their official website.

\subsection{Evaluation on Synthetic Data}
\noindent

\begin{table*}[!t]
\scriptsize
\center
\begin{center}
\caption{Quantitative comparison on various datasets. The \textcolor{gray}{gray} results indicate unfair comparisons. (The degradation models of ZSSR and RCAN both assume bicubic downsampling, while ZSSR-NB operates in a non-blind setting with access to the ground truth of the degradation kernel.) 
The best results of fair method comparison are highlighted in \textbf{bold}.}
\label{tab:si_psnr}
\begin{tabularx}{\linewidth}{X|c|c|c|c|c|c}

\toprule
\multirow{2}*{Method}  & Scale  & Set5 & Set14 &  Urban100 &   Manga109 &  DIV2K100 \\
\cline{3-7}\rule{0pt}{3ex}
~ &  Factor & PSNR/SSIM & PSNR/SSIM &  PSNR/SSIM &    PSNR/SSIM &  PSNR/SSIM \\
\midrule

RCAN~\cite{zhang2018image} & $\times$2 & \textcolor{gray}{26.75} / \textcolor{gray}{0.81} & 
\textcolor{gray}{24.77} / \textcolor{gray}{0.70} & 
\textcolor{gray}{22.46} / \textcolor{gray}{0.66} & 
\textcolor{gray}{23.56} / \textcolor{gray}{0.78} & 
\textcolor{gray}{26.05} / \textcolor{gray}{0.74} \\                                          

ZSSR~\cite{shocher2018zero} & $\times$2 & \textcolor{gray}{26.72 / 0.81} & \textcolor{gray}{24.46 / 0.70} & \textcolor{gray}{22.47 / 0.66} & \textcolor{gray}{23.53 / 0.78} & \textcolor{gray}{26.03 / 0.74} \\
 
ZSSR-NB~\cite{shocher2018zero} & $\times$2 & \textcolor{gray}{33.40 / 0.91} & \textcolor{gray}{30.08 / 0.85} & \textcolor{gray}{27.67 / 0.83} & \textcolor{gray}{32.19 / 0.91} & \textcolor{gray}{31.24 / 0.87} \\
 
DoubleDIP~\cite{ren2020neural} & $\times$2 & 18.57 / 0.48 & 18.90 / 0.44 & 17.18 / 0.41 & 18.44 / 0.58 & 20.40 / 0.50 \\

DIPFKP~\cite{liang2021flow} & $\times$2 & 28.76 / 0.86 & 26.38 / 0.75 & 24.60 / 0.72 & 27.75 / 0.85 & 27.19 / 0.75 \\  

BSRDM~\cite{yue2022blind} & $\times$2 & 32.76 / 0.91 & 28.65 / 0.81 & 25.46 / 0.76 & 28.49 / 0.87 & 28.32 / 0.78 \\

Our & $\times$2 & \textbf{33.51} / \textbf{0.92} & \textbf{29.61} / \textbf{0.83} & \textbf{26.40} / \textbf{0.79} & \textbf{29.89} / \textbf{0.89} & \textbf{29.46} / \textbf{0.82} \\

\midrule %
RCAN~\cite{zhang2018image}       & $\times$3 & \textcolor{gray}{23.12 / 0.68} & \textcolor{gray}{21.71 / 0.55} & \textcolor{gray}{19.71 / 0.51}  & \textcolor{gray}{20.13 / 0.64} & \textcolor{gray}{23.05 / 0.61}
 \\
ZSSR~\cite{shocher2018zero}      & $\times$3 & \textcolor{gray}{23.02 / 0.68} & \textcolor{gray}{21.54 / 0.55} & \textcolor{gray}{19.93 / 0.52}  & \textcolor{gray}{20.21 / 0.65} & \textcolor{gray}{23.14 / 0.62}
 \\
ZSSR-NB~\cite{shocher2018zero}  & $\times$3 & \textcolor{gray}{26.88 / 0.78} & \textcolor{gray}{24.19 / 0.66} & \textcolor{gray}{22.43 / 0.63}  & \textcolor{gray}{24.05 / 0.75} & \textcolor{gray}{26.11 / 0.71}
 \\
% SwinIR~\cite{liang2021swinir}  & $\times$3 & 23.11 / 0.69 & 21.68 / 0.62 & 19.64 / 0.71  & 20.10 / 0.76 & 23.02 / 0.84
%  \\
DoubleDIP~\cite{ren2020neural}   & $\times$3 & 17.39 / 0.42 & 18.22 / 0.41 & 16.96 / 0.40  & 18.19 / 0.56 & 20.10 / 0.50
  \\
DIPFKP~\cite{liang2021flow}      & $\times$3 & 25.47 / 0.82 & 25.06 / 0.72 & 23.51 / 0.69  & 26.54 / 0.83 & 26.16 / 0.72
  \\
BSRDM~\cite{yue2022blind}        & $\times$3 & 30.96 / 0.88 & 27.67 / 0.77 & 24.66 / 0.72  & 27.44 / 0.85 & 27.67 / 0.76
  \\
Our                              & $\times$3 & \textbf{31.37} / \textbf{0.89} & \textbf{28.14} / \textbf{0.79} & \textbf{25.26} / \textbf{0.75}  & \textbf{28.50} / \textbf{0.87} & \textbf{28.51} / \textbf{0.79} \\

\midrule %
RCAN~\cite{zhang2018image}       & $\times$4 & \textcolor{gray}{20.59 / 0.57} & \textcolor{gray}{19.91 / 0.47} & \textcolor{gray}{17.82 / 0.41}  & \textcolor{gray}{17.90 / 0.55} & \textcolor{gray}{21.18 / 0.54}
 \\
 
ZSSR~\cite{shocher2018zero}      & $\times$4 & \textcolor{gray}{20.63 / 0.57} & \textcolor{gray}{20.02 / 0.48} & \textcolor{gray}{18.39 / 0.43}  & \textcolor{gray}{18.29 / 0.56} & \textcolor{gray}{21.45 / 0.55}
 \\
 
ZSSR-NB~\cite{shocher2018zero}  & $\times$4 & \textcolor{gray}{23.24 / 0.62} & \textcolor{gray}{23.19 / 0.60} & \textcolor{gray}{21.89 / 0.59}  & \textcolor{gray}{23.60 / 0.72} & \textcolor{gray}{25.81 / 0.69}
 \\
 
% SwinIR~\cite{liang2021swinir}  & $\times$4 & 20.49 / 0.56 & 19.81 / 0.50 & 17.69 / 0.54  & 17.82 / 0.60 & 21.13 / 0.72
%  \\
DoubleDIP~\cite{ren2020neural}   & $\times$4 & 17.53 / 0.43 & 18.24 / 0.41 & 17.25 / 0.40  & 17.49 / 0.53 & 20.05 / 0.50
  \\

DIPFKP~\cite{liang2021flow}      & $\times$4 & 24.32 / 0.81 & 23.93 / 0.68 & 22.56 / 0.66  & 24.86 / 0.79 & 25.16 / 0.70
  \\

BSRDM~\cite{yue2022blind}        & $\times$4 & 29.02 / 0.85 & 26.63 / 0.73 & 23.47 / 0.68  & 25.77 / 0.81 & 26.74 / 0.72
  \\

Our                              & $\times$4 & \textbf{29.29} / \textbf{0.85} & \textbf{26.81} / \textbf{0.73} & \textbf{23.90} / \textbf{0.70}  & \textbf{26.44} / \textbf{0.83} & \textbf{27.34} / \textbf{0.75}  \\

\bottomrule
\end{tabularx}
\end{center}
\end{table*}

%------------------------------------------------------------------------
\begin{figure*}[ttp]
    \centering
  \setlength{\tabcolsep}{0.1mm}
  \begin{tabular}{ccccccc}
\includegraphics[width=0.14\linewidth]{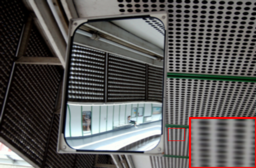}&
\includegraphics[width=0.14\linewidth]{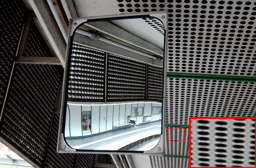}&
\includegraphics[width=0.14\linewidth]{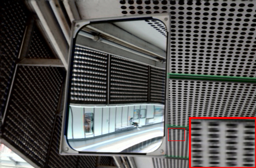}&
\includegraphics[width=0.14\linewidth]{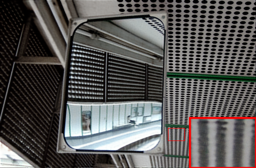}&
\includegraphics[width=0.14\linewidth]{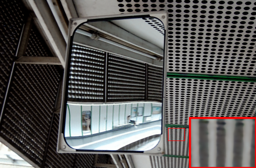}&
\includegraphics[width=0.14\linewidth]{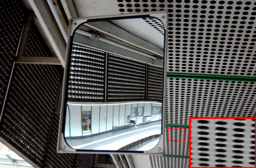}&
\includegraphics[width=0.14\linewidth]{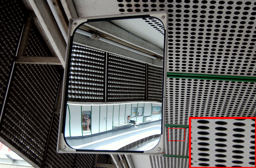}\\
\includegraphics[width=0.14\linewidth]{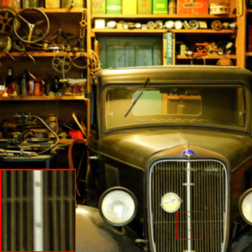}&
\includegraphics[width=0.14\linewidth]{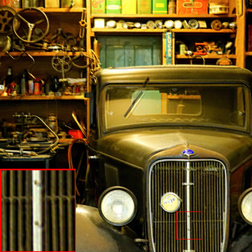}&
\includegraphics[width=0.14\linewidth]{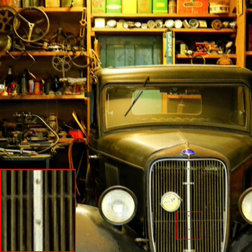}&
\includegraphics[width=0.14\linewidth]{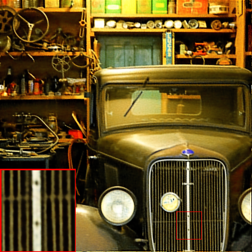}&
\includegraphics[width=0.14\linewidth]{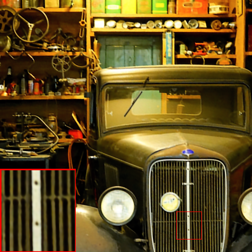}&
\includegraphics[width=0.14\linewidth]{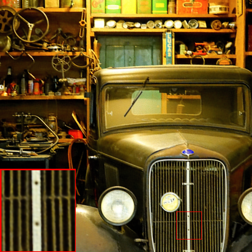}&
\includegraphics[width=0.14\linewidth]{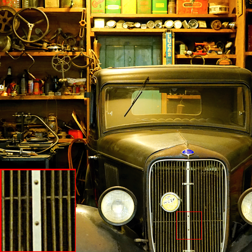}\\
\includegraphics[width=0.14\linewidth]{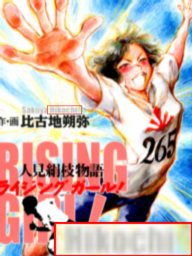}&
\includegraphics[width=0.14\linewidth]{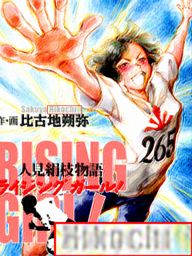}&
\includegraphics[width=0.14\linewidth]{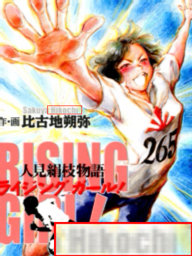}&
\includegraphics[width=0.14\linewidth]{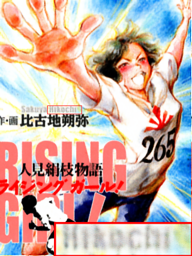}&
\includegraphics[width=0.14\linewidth]{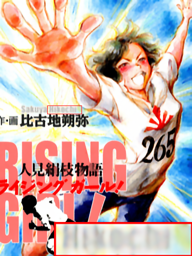}&
\includegraphics[width=0.14\linewidth]{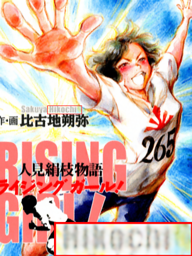}&
\includegraphics[width=0.14\linewidth]{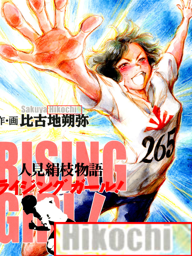}\\
Bilinear & ZSSR-NB & DoubleDIP & DIPFKP & BSRDM & Our & GT \\
  \end{tabular}
    \caption{Visualization of the synthetic images. Each row, from top to bottom, contains super resolved images from Urban100, DIV2K100 and Manga109, under the scale factor of 2, 3 and 4, respectively.}
    \label{fig:vis_testdata}
\end{figure*}
%------------------------------------------------------------------------

We adopt five widely-used datasets to synthesize degradation images for evaluation, including Set5~\cite{set5}, Set14~\cite{set14}, Urban100~\cite{urban100}, Manga109~\cite{manga109} and DIV2K100~\cite{div2k100}. 
The LR image is first randomly blurred by six degradation kernels, and then downsampled with scale factors 2, 3, and 4 respectively. 
In addition, 1\% Gaussian noise is also added to the LR image to better simulate degradation images in real scenes.
The six degradation kernels used are sourced from BSRDM, which includes two isotropic Gaussian kernels with different widths and four isotropic Gaussian kernels.
To address GPU memory limitations, all images larger than 1024 $\times$ 1024 are cropped below the center.
The results are evaluated by the peak signal-to-noise ratio (PSNR, measured in dB) and structural similarity index (SSIM) \cite{wang2004image}. 
PSNR and SSIM are evaluated on the Y channel of the transformed YCbCr color space using MATLAB.
The PSNR and SSIM of different methods under different datasets are shown in \cref{tab:si_psnr}.

\begin{figure}[t]
 \begin{center}
   \begin{minipage}{0.55\linewidth}
    \centering
    \begin{tabular}{c}
      \includegraphics[width=0.98\linewidth]{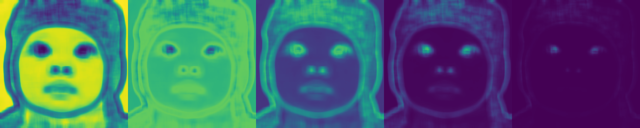}\\
      \includegraphics[width=0.98\linewidth]{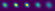}\\
      (a)\\
    \end{tabular}
  \end{minipage}
  \begin{minipage}{0.11\linewidth}
    \centering
    \begin{tabular}{c}
      \includegraphics[width=0.93\linewidth]{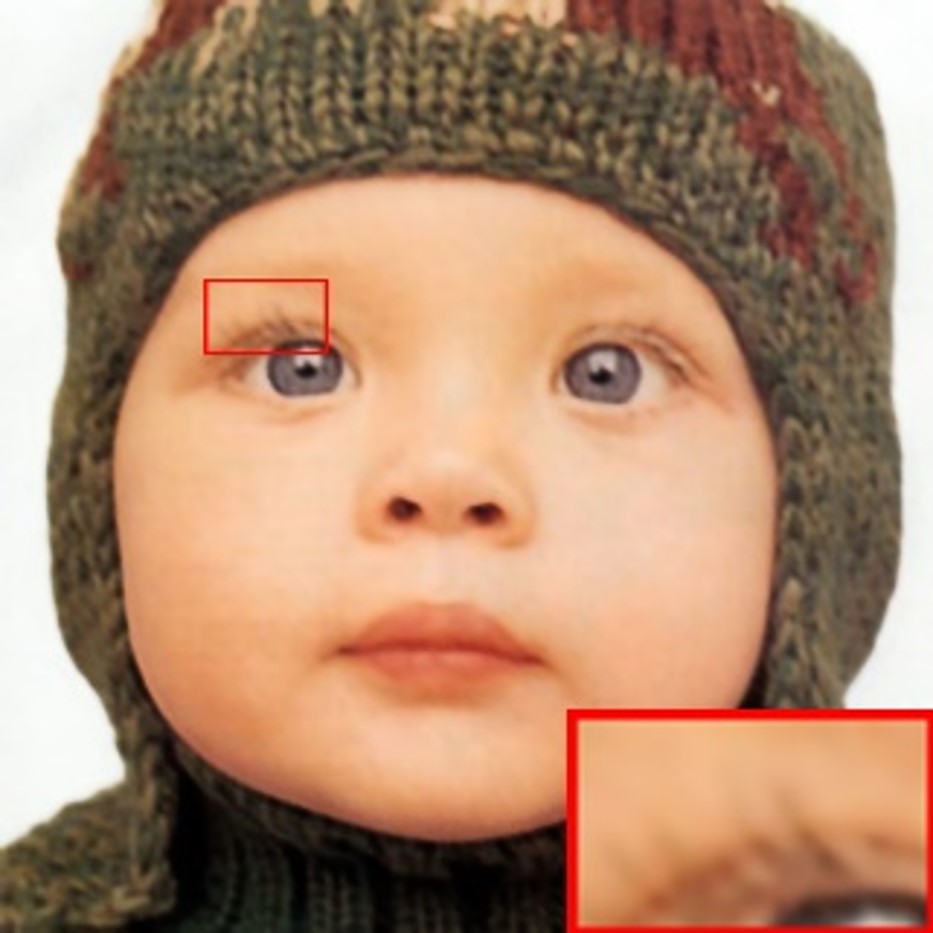}\\
      \includegraphics[width=0.93\linewidth]{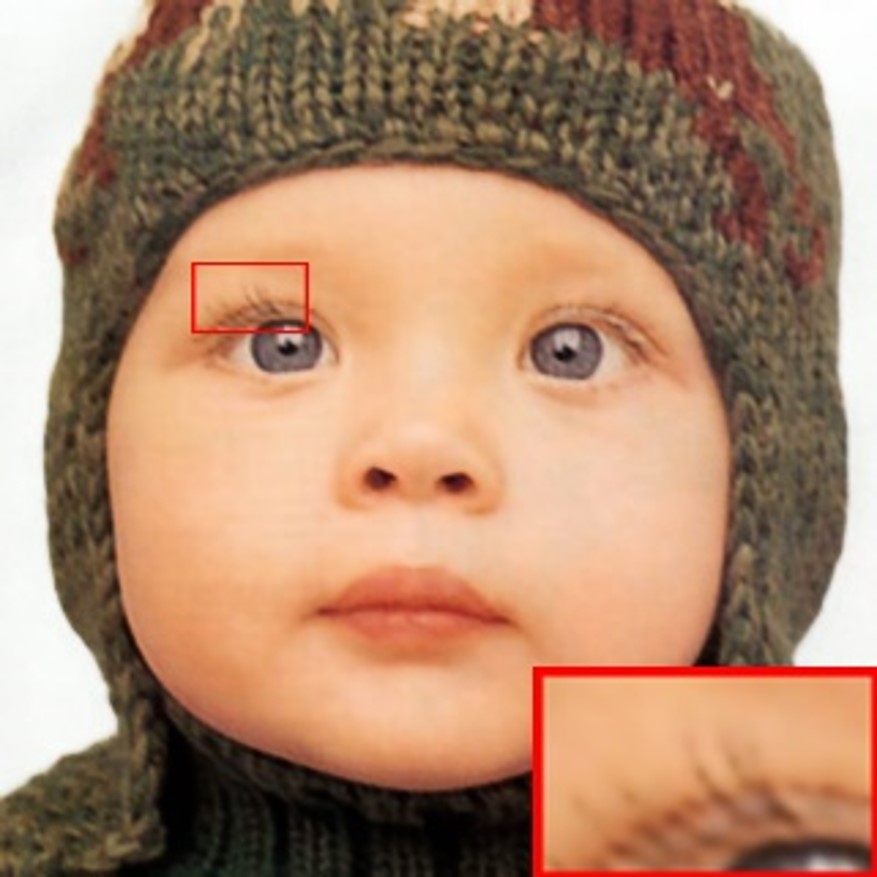}\\
      (b)\\
    \end{tabular}
  \end{minipage}
  \hfill
  \begin{minipage}{0.31\linewidth}
    \centering
    \begin{tabular}{c}
      \includegraphics[width=0.97\linewidth]{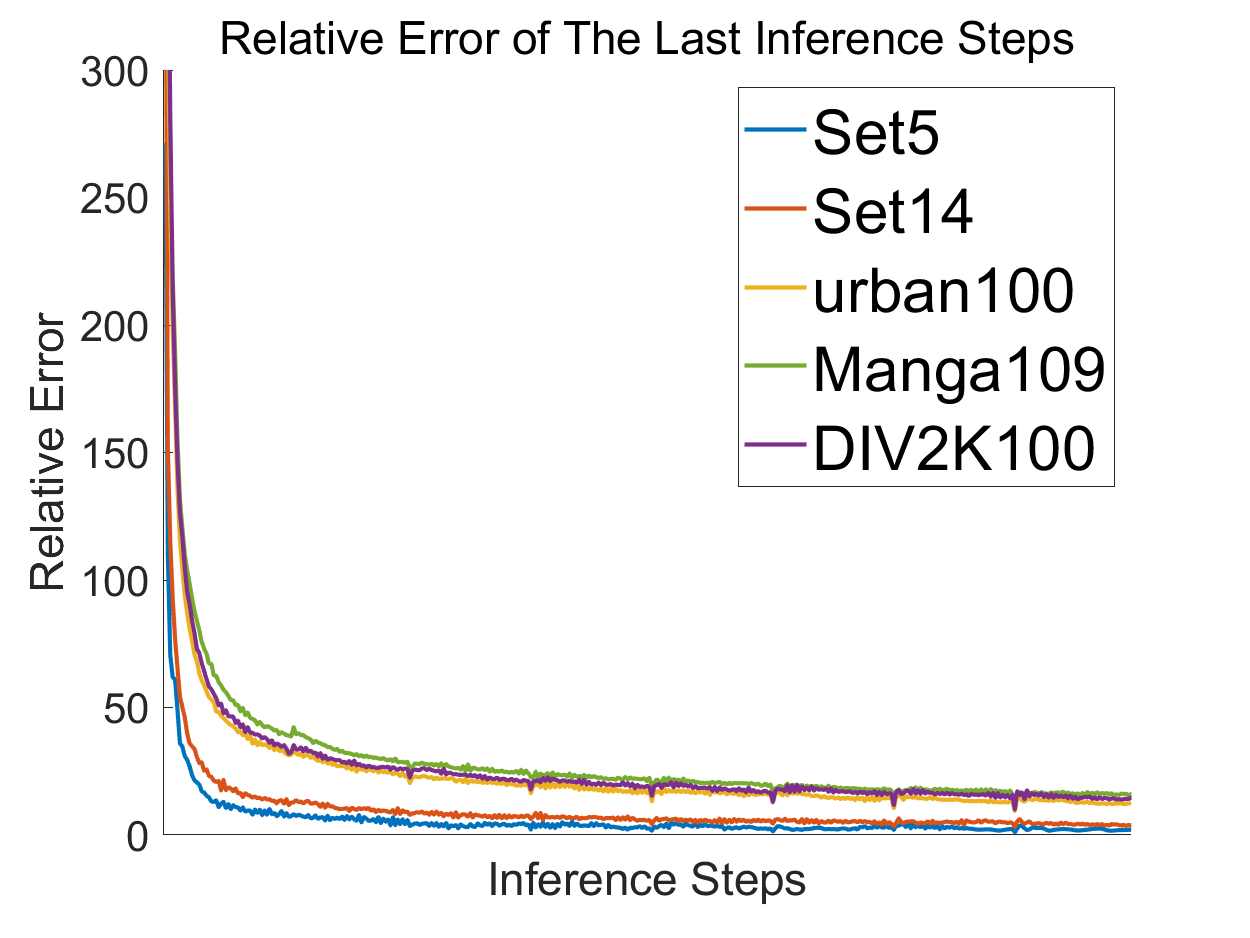}\\
      (c)\\
    \end{tabular}
  \end{minipage}
 \end{center}
 \caption{(a) Visualization of the coefficient matrices and the corresponding learned atom degradation kernels of the image Baby under the scale factor of 2 in Set 5. (b) Visualization of the synthetic images with the spatially-invariant degradation model (top) and spatially-variant degradation model (bottom). (c) Convergence curves.}
 \label{fig:sgatt}
\end{figure}

It is evident that our proposed method SVDSR outperforms the SOTA method BSRDM in 2$\times$ SR,, and also has certain advantages in 4$\times$SR.
This is because compared with the spatially-invariant degradation model, the proposed spatially-variant degradation model fully utilizes the physical information in the image, but at 4$\times$SR, due to the excessive loss of texture information in the image, the superiority of the SVDSR decreases to a certain extent.
The performance of Data-driven SR methods RCAN and ZSSR significantly degrades when the degradation kernel deviates from their bicubic degradation assumptions.
In addition, SVDSR remains competitive with the non-blind methods ZSSR-NB, 
which strongly demonstrates the effectiveness of the proposed spatially-variant degradation model in estimating the degradation kernel.
The superior performance on the natural image dataset DIV2K100 and the cartoon image dataset Manga109 shows the good adaptability of SVDSR on different images.

\cref{fig:vis_testdata} compares the visual effects with varying scale factors.
It is obvious that our method can achieve favorable results in photos under different scenarios, especially in areas with complex details.
This is attributed to the substantial differences in degradation  kernel information between flat areas and densely textured areas in the image. 
Our proposed SVDSR leverages each pixel in the image to estimate its own degradation  kernel, utilizing the texture information of the region in which the pixel is located.
This approach enhances our ability to model image degradation, resulting in superior performance in BISR.
\cref{fig:sgatt}(a) shows the derived coefficient matrices and the corresponding learned atom degradation kernels exported from the baby image.
It can be seen that the derived coefficient matrices well reflect the texture information of the image and the detail reconstruction effect under the spatially-variant model in \cref{fig:sgatt}(b)  is remarkable.
Moreover, there are significant differences in the shape of atom degradation kernels in different regions, which is exactly the same phenomenon observed in LARPAR\cite{li2020lapar}.
\cref{fig:sgatt}(c) demonstrates the convergence on different datasets.

\subsection{Evaluation on Real Data}
\noindent

\begin{figure*}[t]
  \centering
  \setlength{\tabcolsep}{0.2mm}
  \begin{tabular}{ccccccc}
  \includegraphics[width=0.139\linewidth]{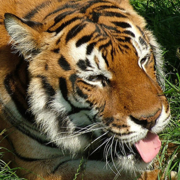}&
\includegraphics[width=0.139\linewidth]{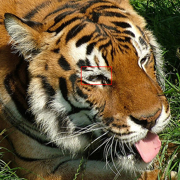}&
\includegraphics[width=0.139\linewidth]{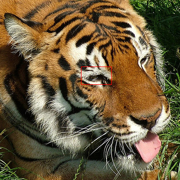}&
\includegraphics[width=0.139\linewidth]{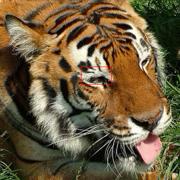}&
\includegraphics[width=0.139\linewidth]{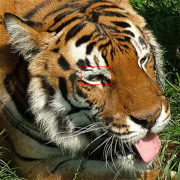}&
\includegraphics[width=0.139\linewidth]{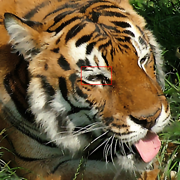}&
\includegraphics[width=0.139\linewidth]{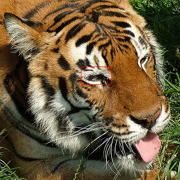}\\
\includegraphics[width=0.139\linewidth]{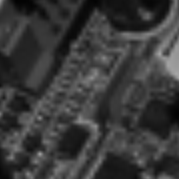} &
\includegraphics[width=0.139\linewidth]{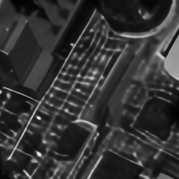}&
\includegraphics[width=0.139\linewidth]{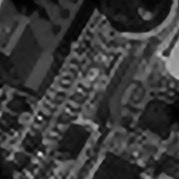}&
\includegraphics[width=0.139\linewidth]{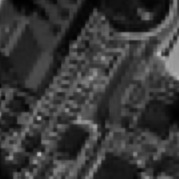}&
\includegraphics[width=0.139\linewidth]{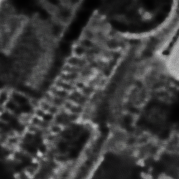}&
\includegraphics[width=0.139\linewidth]{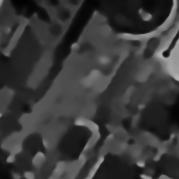}&
\includegraphics[width=0.139\linewidth]{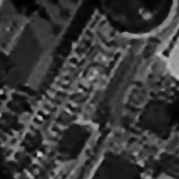}\\
Bilinear &  RCAN & ZSSR  & DoubleDIP & DIPFKP & BSRDM & Our  \\
\end{tabular}
\caption{Visualization of the RealSRSet under scale factor of 2(top) and 4(bottom).}
\label{fig:vis_real_image}
\end{figure*}

To further demonstrate the applicability of our method to real images, we evaluated data-driven and dataset-free methods using a small dataset called RealSRSet \cite{zhang2021designing}, which consists of 20 real images. 
As ground truth for real-world images is unavailable, we performed visual comparisons only on the HR images generated by different methods.
A subset of estimated HR images is depicted in \cref{fig:vis_real_image}. 
It is evident that our proposed method outperforms other dataset-free methods significantly in terms of detail restoration and is comparable to data-driven methods.
Given that the SOTA method BSRDM also incorporates denoising during BISR, detailed areas such as the tiger's hair exhibit varying degrees of blurring in BISR for real images.

\subsection{Comparison on Model Size and Runtime}
\noindent

We further compare different models in terms of running time and memory. 
The comparison results are shown in \cref{Tab:ablation_time}. 
All tests were conducted on the GeForce RTX 3090 GPU. 
We test the running time of super-resolution with scale factors of 2, 3, and 4 for 256 $\times$ 256 images. 
For the sake of fairness, only four methods based on traditional degradation models are considered, namely ZSSR, DoubleDIP, DIPFKP, and BSRDM. 
It can be seen that even though our proposed SVDSR estimates its own degradation  kernel for each pixel, its time and memory consumption do not exceed BSRDM significantly. 
This suggests that our model has greater potential for practical utilization than these methods in real-world scenarios.

\subsection{Ablation Studies and Analysis}
\noindent

\noindent {\bf Ablation Studies\ } 
To assess the efficiency of each module in our proposed method, we construct three baselines on Set 14 under the scale factor
of 2 to analyze the effectiveness of the innovative likelihood and image prior in our proposed SVDSR.
\begin{itemize}[leftmargin=1em, itemindent=0em]
\setlength{\itemsep}{0pt}
\item Case1: Likelihood only contains the spatial domain, and the frequency domain term is removed.
\item Case2: The results of the encoder in the generation network is directly concatenated with the input of the corresponding decoder, instead of using FFT(Fast Fourier Transform) and IFFT(Inverse Fast Fourier Transform).
\item Case3: Instance layer in the network is removed to verify the Instance layer's alleviating effect on the overfitting phenomenon of the generation network.
\end{itemize}

As shown in \cref{Tab:ablation_case}, when the skip structure of the generation net and the likelihood function in the frequency domain are removed, the PSNR index of the data results is reduced to some extent. After removing the instance layer used to alleviate the overfitting phenomenon of the generation net, the results of our model are significantly diminished. This study demonstrates that our proposed novel likelihood function, and image prior all contribute to improving the performance of our method and are applicable to other network architectures.

\noindent {\bf Analysis of The Atom Kernel Dictionary}
We investigated the impact of the number of atom degradation kernels on our proposed spatially-variant degradation model.
We found that the optimal number of atom degradation kernel varies for different images, but it is usually a single digit.
This indicates that the 72 pre-defined atom kernel used in the previous method\cite{li2020lapar,zhou2023learning} are redundant.
Due to the good consistency of image characteristics in the set5, the experiment was conducted on set 5 under the scale factor of 2.
\cref{Tab:ablation_nd} shows the effect of our model with different number $N_\mathcal{D}$ of atom degradation kernels and non-blind setting. 
When $N_\mathcal{D}$=1, our fuzzy degradation model is equivalent to the spatially-invariant degradation model. 
The effectiveness of our model improves initially as \(N_\mathcal{D}\) increases.
However, as \(N_\mathcal{D}\) exceeds 5, the performance of the proposed model starts to decline. 
\cref{tab:table_para} demonstrates the impact of hyperparameters.
We follow BSRDM for hyperparameter settings for fairness. 
We can even achieve better results in other parameter settings, \emph{e.g.}, $\sigma_f$ = 20, $\sigma_x$=25.

\begin{table}[t]
\centering
\hfill
\begin{minipage}{0.74\linewidth}
\caption{Comparison of different methods on model size, runtime and iterations for 2$\times$ BISR..}
\label{Tab:ablation_time}
\centering
\resizebox{\linewidth}{!}{
\begin{tabular}{c|c|c|c|c|c}
\hline
Methods & ZSSR & DoubleDIP & DIPFKP & BSRDM & SVDSR \\
\hline
Time & 63s & 220s & 68s & 30s & 33s \\
\hline
Parameters & 225k & 2396k & 2396k & 762k & 850k \\
\hline
Iterations & 3000 & 2500 & 1000 & 5000 & 5000 \\
\hline
\end{tabular}
}
\end{minipage}
\begin{minipage}{0.25\linewidth}
\caption{PSNR of different case.}
\label{Tab:ablation_case}
\centering
\resizebox{\linewidth}{!}{
\begin{tabular}{c|c}
\hline
Case 1 & 29.61 \\
\hline
Case 2 & 29.52 \\
\hline
Case 3 & 29.28 \\
\hline
\end{tabular}
}
% }
\end{minipage}
\end{table}

\begin{table}[t]
\scriptsize
\centering
\hfill
\begin{minipage}{0.71\linewidth}
\caption{The impact of the number of atom blur kernels(PSNR).}
\label{Tab:ablation_nd}
\centering
\resizebox{\linewidth}{!}{
\begin{tabular}{c|c|c|c|c|c|c}
\hline
Number & $N_\mathcal{D}$=1 & $N_\mathcal{D}$=3 & $N_\mathcal{D}$=5 & $N_\mathcal{D}$=7 & $N_\mathcal{D}$=9 & NB \\
\hline
Set5 & 33.19 & 33.29 & 33.51 & 33.43 & 33.27 & 34.44 \\
Set14 & 29.13 & 29.40 & 29.61 & 29.58 & 29.59 & 30.32\\
Urban100 & 25.90 & 26.11 & 26.40 & 26.21 & 26.20 & 27.14\\
Manga109 & 29.23 & 29.53 & 29.89 & 29.67 & 29.74 & 31.13 \\
DIV2K100 & 29.05 & 29.30 & 29.46 & 29.52 & 29.54 & 30.06\\
\hline
\end{tabular}
}
\end{minipage}
\begin{minipage}{0.28\linewidth}
\caption{The impact of  hyperparameter.}
\label{tab:table_para}
\centering
\resizebox{\linewidth}{!}{
\begin{tabular}{c|c|c|c}
\hline
$\sigma_x$ &  & $\sigma_y$ &  \\
\hline
0.25 & 27.06 & 0.1 & 29.56\\
25 & 29.74 & 10 & 28.30\\
\hline
$\sigma_{\gamma}$ &  & $\sigma_f$ &  \\
\hline
0.15 & 29.53 & 0.2 & 29.39 \\
15 & 29.60 & 20 & 29.63 \\
\hline
\end{tabular}
}
\end{minipage}
\end{table}

\section{Conclusion and Limitations}
\label{sec:conclu}
\noindent
This paper introduces a fuzzy degradation model to deal with the ambiguity of dataset-free Blind Image Super-Resolution (BISR) solutions in different regions. We present the degradation  kernel of each pixel as a linear combination of a learnable dictionary consisting of a small number of fuzzy atom kernels. The coefficients of these atom degradation  kernels are derived using membership functions, allowing us to capture the uncertainty and variability in the degradation process. We propose a Probabilistic BISR framework with customized likelihood and prior terms. The Monte Carlo EM algorithm, which iteratively estimates the degradation  kernels of each pixel while considering the probabilistic nature of the problem is employed. Extensive experiments show the superiority of our method.
Color distortion is a challenge faced by most SR methods due to strong noise in several images. 
Similarly, our method also encounters this problem, and our future goal is to achieve better denoising while preserving more details.

\section*{Acknowledgements}
This work was supported by the National Natural Science Foundation of China (Grant No. 62101191), Shanghai Natural Science Foundation (Grant No.  21ZR1420800), and the Science and Technology Commission of Shanghai Municipality (Grant No. 22DZ2229004).

\bibliographystyle{splncs04}
\bibliography{main}
\end{document}